\documentclass[aps,pra,reprint,superscriptaddress,flushbottom,longbibliography]{revtex4-1}

\usepackage{general_latex}

\begin{document}

\title{Amplification of the quantum superposition macroscopicity of a flux qubit \\ by a 
magnetized Bose gas}
\author{T.\,J.~Volkoff}
\email{volkoff@snu.ac.kr}
\author{Uwe R. Fischer}
\affiliation{Department of Physics and Astronomy, Center for Theoretical Physics, Seoul National University, Seoul 08826, Korea}

\begin{abstract}
We calculate a measure of superposition macroscopicity $\mathcal{M}$ for a superposition of screening current states in a superconducting flux qubit (SFQ), by relating $\mathcal{M}$ to the action of an instanton trajectory connecting the potential wells of the flux qubit. When a magnetized Bose-Einstein condensed (BEC) gas containing $N_B\sim \ord (10^6)$ atoms is brought into a $\mathcal{O}(1)$ $\mu\text{m}$ proximity of the flux qubit in an experimentally realistic geometry, 
we demonstrate the appearance of a two- to five-fold amplification of $\mathcal{M}$ over the bare 
value without the BEC, by calculating the instantion trajectory action from the microscopically derived effective flux Lagrangian of a hybrid quantum system composed of the flux qubit and a spin-$F$ atomic Bose gas.
Exploiting the connection between $\mathcal M$ and the maximal metrological usefulness of a multimode superposition state, we show that amplification of $\mathcal{M}$ in the ground state of the hybrid system is equivalent to a decrease in the quantum Cram\'{e}r-Rao bound for estimation of an externally applied flux. Our result therefore demonstrates the increased usefulness of the BEC--SFQ hybrid system as a sensor of ultraweak magnetic fields below the standard quantum limit.
\end{abstract}
\maketitle

\section{\label{sec:intro}Introduction}
The development of hybrid quantum devices for quantum information processing
grows out of the desire to achieve disparate and seemingly mutually exclusive 
goals: quantum information should be both long-lived and immune to decoherence, as well as easy to retrieve and manipulate \cite{Wallquist,Kurizki}.
These goals have been intensely pursued in hybrid quantum devices at the interface of solid-state physics and atomic physics, including  
superconductors interacting with other quantum systems (see, e.g., 
Refs.\cite{RevModPhys.85.623,PhysRevLett.105.210501,Zhu,taddei,PhysRevA.89.010301,PhysRevLett.100.170501,munrorobust}), inter alia, with ultracold atomic Bose gases 
\cite{PhysRevLett.103.043603,PhysRevA.87.052303,PhysRevLett.111.240504,Bernon,weisstubingen}.

In this paper, we reveal two related facets of a hybrid quantum system consisting of a superconducting flux qubit (SFQ) and magnetized Bose-Einstein condensate (BEC): (i) its increased usefulness {over a non-hybrid SFQ} for external magnetic field sensing, and (ii) the larger scale on which it exhibits quantum mechanical behavior as quantified by an appropriate measure $\mathcal{M}$. For $J$ a finite index set having $\vert J\vert$ elements, the macroscopicity $\mathcal{M}$ of a superposition $\ket{\Psi}$ of product states in $(\mathbb{C}^{2})^{\otimes \vert J \vert}$ is defined as the maximal variance in $\ket{\Psi}$ \begin{equation} \mathcal{M}:= {1\over \vert J \vert} \max_{H} \left(\langle \Psi \vert H^{2} \vert \Psi \rangle -\langle \Psi \vert H \vert \Psi \rangle^{2} \right)\label{eqn:macrodef}\end{equation} over all 1-local observables $H$, i.e., having the form $H = \sum_{j \in J} T^{(j)} \otimes \mathbb{I}^{J\setminus \lbrace j \rbrace}$ with $\Vert T^{(j)} \Vert = 1$ \cite{dur}. Specifically, we have in mind the case for which $J$ is a subset of momentum space containing the Fermi momentum of a superconducting metal and $\ket{\Psi} \propto \ket{\circlearrowleft} + \ket{\circlearrowright}$ is a superposition of screening currents in the SFQ. Several measures of pure state macroscopicity for finite dimensional quantum systems have been shown to be equivalent to $\mathcal{M}$ \cite{frowislink}. Whereas highly correlated quantum superpositions, e.g., Greenberger-Horne-Zeilinger states, entangled coherent states, have been shown to exhibit $\mathcal{M}$ scaling as $\mathcal{O}(\vert J \vert)$ \cite{shimizu,marquardt,whaleyjan,volkoff,hornbergermech,cirac,jeongrev,frowislink,hornbergernature}, certain two-component superpositions occuring in non-critical degenerate matter, such as the ground state flux superposition of a superconducting flux qubit (SFQ), exhibit much lower macroscopicity values \cite{korsbakken,hornbergermech,leggettrev}. Motivation for understanding how the superposition macroscopicity $\mathcal{M}$ of SFQ can be increased arises due the close relationship between $\mathcal{M}$ and optimal usefulness of the superposition state as a probe in an appropriate quantum metrology protocol (see Eq.\,(\ref{eqn:cramraomacro}) below). In addition to their prominence as information carriers in quantum computers \cite{clarke}, SFQs have been proposed as highly sensitive magnetometers \cite{bal,shnyrkov2}, with theoretically greater flux sensitivity compared to an rf-SQUID device \cite{greenberg}.

\begin{figure}[hbt]
\includegraphics[scale=0.42]{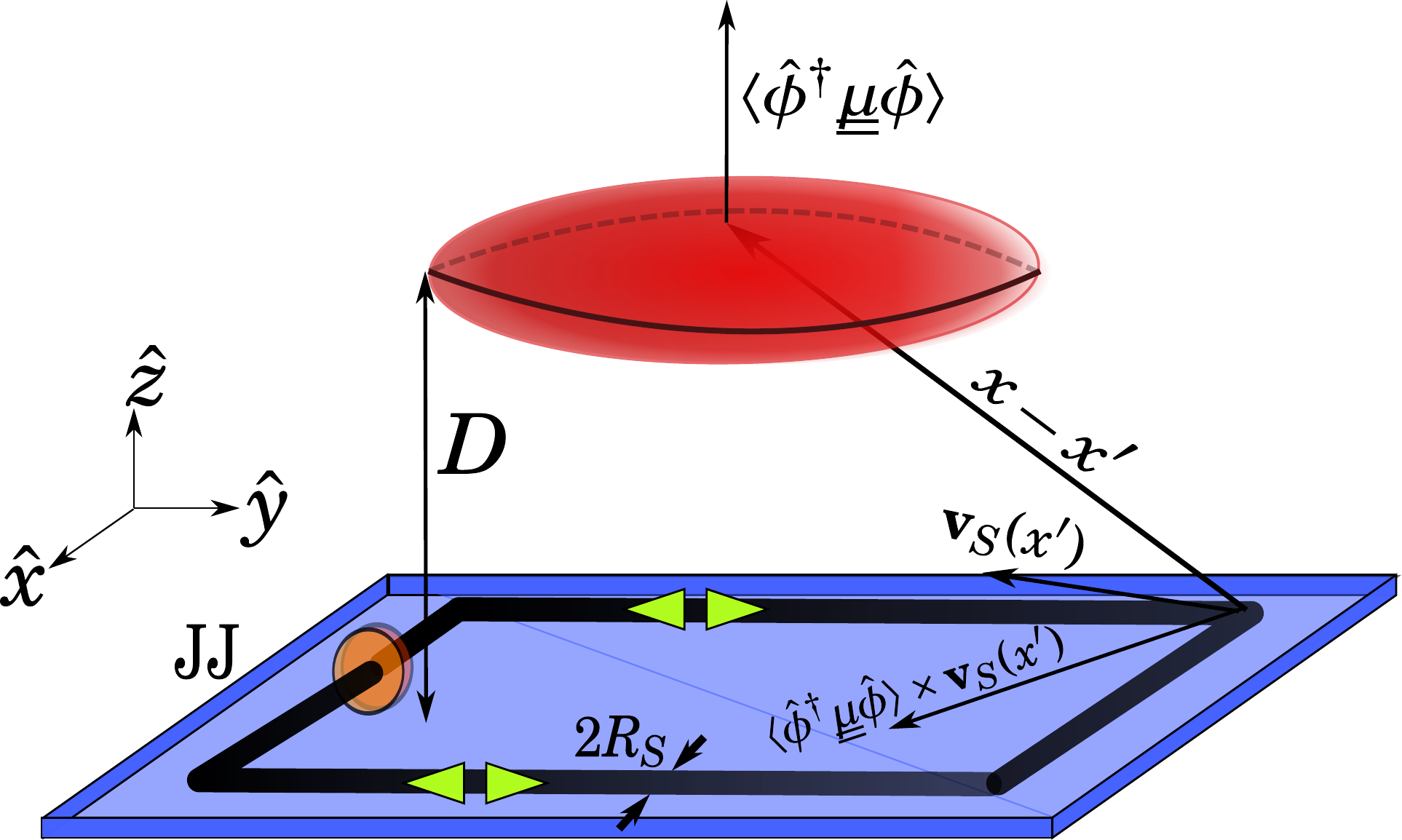}
\caption{Schematic of the geometry and relevant dynamical quantities of the proposed hybrid flux qubit, where the oblate BEC (red) is positioned at distance $D$ above the superconducting loop circuit of wire thickness $2R_S$ (black). The tunnel junction (orange) is labeled JJ and the ground state screening current superposition is shown by the green arrows; ${\bm v}_S$ is the superflow velocity. The local magnetic moment density of the atoms  
$\langle \hat{{\bm\phi}}^{\dagger} \mathuu{{\bm \mu} }\hat{\bm \phi} \rangle$ contains 
$\displaystyle \mathuu{{\bm \mu} } :=  g_{f}\mu_{B}(\mathuu{\sigma_{x}},\mathuu{\sigma_{y}},\mathuu{\sigma_{z}})$ and 
the spin-$F$ field operator $\hat{\bm \phi}$.}
\label{fig:sapphireembedding}
\end{figure}

The main result of the present work is that the macroscopic quantum character of a ground state screening current superposition in a SFQ, quantified by the measure $\mathcal{M}$, can be amplified by coupling the electric current of the superconductor, on the microscopic interaction level, to the local magnetic moment of a proximal spinor Bose gas (Fig.\,\ref{fig:sapphireembedding}). Our proof proceeds according to the following schematic logic: (i) $\mathcal{M}$ increases when the value of $-\ln\vert \langle \circlearrowleft \vert \circlearrowright \rangle \vert$ increases (Sec. \ref{sec:macrosec}); (ii) The quantity $-\ln\vert \langle\circlearrowleft \vert \circlearrowright \rangle \vert$ increases linearly with the Euclidean action of an instanton propagating between the potential wells of the SFQ (Sec. \ref{sec:macrosec}); (iii) The Euclidean action of an instanton can be amplified by the proximal spinor Bose gas (Sec. \ref{sec:effect}). In Sec. \ref{sec:qcr},  we relate the amplification of $\mathcal{M}$ to an increase in usefulness of the hybrid system for quantum magnetometry over a SFQ alone.

Physically, the amplification of $\mathcal{M}$ is a consequence of the large renormalization of the inductance of a SFQ in the proximity of a magnetized Bose gas (e.g., a spin-$F$ BEC). In particular, we show that when the externally applied flux is tuned to $\Phi_{0}/2$ so that the SFQ potential has degenerate minima \cite{leshouchesdevoret,shnyrkov1}, the inductance renormalization $\mathcal{K}>0$ appearing in the effective action of the flux variable $S_{\text{eff}}^{(2)}=\int_{0}^{\beta\hbar}d\tau [C({d\Phi \over d\tau})^{2} + V(\Phi)]$ where
\begin{equation}
V(\Phi)= {I_{c}\Phi_{0} \over 2\pi}\cos  {2\pi\Phi \over \Phi_{0}} + \left({1\over 2L} + \mathcal{K}\right) \Phi^{2}
\label{eqn:fluxactbare}
\end{equation} to second order in perturbation theory, is $\mathcal{O}(1/2L)$ for appropriate geometric arrangements of the hybrid device and an experimentally achievable density of Bose condensed atoms.
In Eq.\,(\ref{eqn:fluxactbare}), $\Phi_{0} = \pi \hbar / e$ is the flux quantum in SI units, $I_{c}$ is the critical current of the Josephson junction, $C$ is the capacitance of the tunnel junction, and $L$ is the self-inductance of the SFQ. This doubling of $\mathcal{K}$ arises due to an approximate cancellation of the small atomic magnetic moments in the Bose gas by the large density of states at the Fermi surface of the superconductor, cf.\,Eqs.\,(\ref{eqn:intermed}) and 
(\ref{eqn:kresult}) below.

\section{\label{sec:macrosec}Macroscopicity of flux qubit screening current superposition state}
We begin by showing that the SFQ ground state $\ket{G}_{SFQ}$ for the model of Eq.\,(\ref{eqn:fluxactbare}) is of the general form \begin{equation}
{1\over \sqrt{\mathcal{N}}}\left( \bigotimes_{j\in J}\ket{\phi_{j}} +  \bigotimes_{j\in J}\ket{\psi_{j}} \right),
\label{eqn:generalsup}
\end{equation}
and calculating $\mathcal{M}$ for this general superposition state. $\ket{G}_{SFQ}$ is an equal weight superposition of screening currents $\ket{G}_{SFQ} := \frac{1}{\sqrt{\mathcal{N}}}(\ket{\circlearrowleft} + \ket{\circlearrowright})$ which can be described microscopically as a superposition of variational Bardeen-Cooper-Schrieffer ground states \cite{bcs} $\ket{\circlearrowleft}$ and $\ket{\circlearrowright}$ corresponding to Cooper pairing at center of mass momenta ${\bm Q}_{L}$ and ${\bm Q}_{R}$ respectively. Using $\ket{0}_{\bm k}$ ($\ket{1}_{\bm k}:= a_{{\bm k},\uparrow}^{\dagger}a_{-{\bm k},\downarrow}^{\dagger}\ket{0}_{\bm k}$) to denote the absence (presence) of an $s$-wave Cooper pair of relative electron momentum $2{\bm k}$, and $\sigma_{j}^{({\bm k})}$ ($j=1,2,3$) to be the Pauli matrices in the $\lbrace \ket{0}_{\bm k},\ket{1}_{\bm k} \rbrace$ subspace, $\ket{G}_{SFQ}$ takes the form of Eq.\,(\ref{eqn:generalsup}) with $\ket{\phi_{\bm k}}:=  \exp (-i\tan^{-1}(v_{\bm k}^{{\bm Q}_{L}}/u_{\bm k}^{{\bm Q}_{L}})\sigma_{y}^{({\bm k})}) \ket{0}_{\bm k}$, $\ket{\psi_{\bm k}}:=  \exp (-i\tan^{-1}(v_{\bm k}^{{\bm Q}_{R}}/u_{\bm k}^{{\bm Q}_{R}})\sigma_{y}^{({\bm k})}) \ket{0}_{\bm k}$. The normalization factor is $\mathcal{N}=(2+2e^{-\lambda})$, where $e^{-\lambda} := \langle \circlearrowright \vert \circlearrowleft \rangle = \prod_{\bm k}u_{\bm k}^{{\bm Q}_{L}}u_{\bm k}^{{\bm Q}_{R}} +v_{\bm k}^{{\bm Q}_{L}}v_{\bm k}^{{\bm Q}_{R}} $. However, for a SFQ with multiple Josephson junctions and exposed to engineered external fields, an accurate calculation of the momenta amplitudes $v_{\bm k}^{{\bm Q}_{R}}$, $v_{\bm k}^{{\bm Q}_{L}}$, from which $\mathcal{M}$ can be inferred exactly, must be obtained from a self-consistent solution of the Bogoliubov equations. In Proposition 1, we calculate $\mathcal{M}$ for the superposition Eq.(\ref{eqn:generalsup}). Subsequently, we show how the value $\mathcal{M}$ for the state $\ket{G}_{\text{SFQ}}$ depends on the physical parameters of the system by relating it to the action of an instanton trajectory connecting the minima of $V(\Phi)$. Later, in Sec. \ref{sec:effect}, we use the same relation to show that an increase of the inductive term in $V(\Phi)$ due to a proximal magnetized Bose gas results in an amplification of $\mathcal{M}$ for the ground state screening current superposition in the SFQ.

The technique used in the proof of Proposition 1 is similar to that used in Ref.\,\cite{volkoffoneparam} to calculate $\mathcal{M}$ for a state of the form $\propto \ket{\phi}^{\otimes \vert J \vert} + U^{\otimes \vert J\vert}\ket{\phi}^{\otimes \vert J \vert}$ for a unitary $U$. For the statement of Proposition 1, let $\lbrace \ket{\phi_{ j}}\rbrace_{{j\in J}}$ and $\lbrace\ket{\psi_{ j}}\rbrace_{j\in J}$ in Eq.(\ref{eqn:generalsup}) be collections of normalized pure states indexed by a finite set $J$ having cardinality $\vert J\vert$ and define $z_{j} := \langle \phi_{j} \vert \psi_{j} \rangle$ for all $j \in J$. 

\textit{Proposition 1}: The normalized superposition state \begin{equation}
\ket{G}:= {1\over \sqrt{2+2\text{Re}\prod_{j}z_{j}}}\left( \bigotimes_{j\in J}\ket{\phi_{j}} + \bigotimes_{j\in J}\ket{\psi_{j}} \right)
\label{eqn:supp1}
\end{equation}
has macroscopicity \begin{equation}
\mathcal{M} =  1+{\sum_{j\neq k}\sqrt{(1-\vert z_{j} \vert^{2})(1-\vert z_{k} \vert^{2})}\over \vert J \vert \left( 1+\text{Re}\prod_{j }z_{j}\right) }.
\label{eqn:mvalue}
\end{equation}

\textit{Proof of Proposition 1}:  Consider the operator $H = \sum_{j \in J} T^{(j)}$ where $T^{(j)} = (\ket{\phi_{j}}\bra{\phi_{j}} - \ket{\psi_{j}}\bra{\psi_{j}})/\sqrt{1-\vert z_{j}\vert^{2}}$. It is clear that $\Vert T^{(j)} \Vert = 1$. Then a simple calculation shows that \begin{eqnarray}
\langle (\Delta H)^{2} \rangle_{\ket{G}}& =&  \vert J \vert + {\sum_{j\neq k  }\sqrt{(1-\vert z_{j} \vert^{2})(1-\vert z_{k} \vert^{2})}\over  1+\text{Re}\prod_{j}z_{j}}
\label{eqn:variance}
\end{eqnarray}
We now show that $H$ exhibits maximal variance over the set of all 1-local observables on $(\mathbb{C}^{2})^{\otimes \vert J \vert}$  having operator norm equal to 1. The proof is by induction: let $J = \lbrace 1,\ldots , N \rbrace$. the base case is to consider the states $\ket{\phi_{j}}$ and $\ket{\psi_{j}}$ in $\mathbb{C}^{2}$ and let $\langle \phi_{j} \vert \psi_{j} \rangle \in \mathbb{R}$ without loss of generality. Form the orthonormal basis $\ket{e_{\pm}^{(j)}} := (\ket{\phi_{j}}\pm \ket{\psi_{j}})/\sqrt{2\pm 2z_{j}}$ and define the Pauli operators $\sigma_{x}^{(j)}$, $\sigma_{y}^{(j)}$,  $\sigma_{z}^{(j)}$ with $\sigma_{z}^{(j)}\ket{e_{\pm}^{(j)}}=\pm \ket{e_{\pm}^{(j)}}$. With $z_{j} \in \mathbb{R}$, the most general norm 1 operator has the form $ T^{(j)}_{\theta} = \sigma_{x}^{(j)} \sin \theta+  \sigma_{z}^{(j)}\cos \theta$. For a single mode, say, mode 1 in $J$, the state in Eq.\,(\ref{eqn:supp1}) becomes $\ket{e_{+}^{(1)}}$ and the maximal variance of $T^{(1)}_{\theta}$ in $\ket{e_{+}^{(1)}}$ is achieved for $\theta = \pi/2$, i.e., $T^{(1)}_{\pi/2} = (\ket{\phi_{1}}\bra{\phi_{1}} - \ket{\psi_{1}}\bra{\psi_{1}})/\sqrt{1-z_{1}^{2}}$. Assume now that the 1-local, norm one operators $H_{M} := \sum_{j=1}^{M}T^{(j)}_{\pi /2}$ maximize the variance in $\ket{G_{M}}:= (\otimes_{j=1}^{M}\ket{\phi_{j}} + \otimes_{j=1}^{M}\ket{\psi_{j}})/ \sqrt{2+2\prod_{j=1}^{M}z_{j}}$ for all $M \in \lbrace 1 , \ldots , N-1 \rbrace$. We calculate the maximum over $\theta$ of the variance of the operator $H_{\theta}:= \mathbb{I}^{J\setminus \lbrace N \rbrace} \otimes T^{(N)}_{\theta} + H_{N-1}\otimes \mathbb{I}^{J\setminus \lbrace 1,\ldots ,N-1 \rbrace} $ in the state of Eq.\,(\ref{eqn:supp1}).

We find that \begin{equation} \langle G \vert H_{\theta}\vert G\rangle = {\prod_{j=1}^{N-1}z_{j}^{2} \over (1+\prod_{j=1}^{N-1}z_{j})^{2}}\cos^{2}\theta \end{equation} and \begin{equation}
\langle G \vert H_{\theta}^{2}\vert G \rangle = \langle \mathbb{I}^{J} + H_{N-1}^{2}\otimes \mathbb{I}^{J\setminus \lbrace 1,\ldots ,N-1 \rbrace} + 2T^{(N)}_{\theta}\otimes H_{N-1} \rangle_{\ket{G}}.
\label{eqn:varsupp}\end{equation} Taking the partial derivative of Eq.\,(\ref{eqn:varsupp}) with respect to $\theta$ produces the condition \begin{equation}
\cot \theta = {\langle \sigma_{z}^{(N)} \otimes H_{N-1} \rangle_{\ket{G}} \over \langle \sigma_{x}^{(N)} \otimes H_{N-1} \rangle_{\ket{G}} }.
\end{equation} Since $\braket{\phi_{N}}{\sigma_{z}^{(N)}\vert \phi_{N}} = \braket{\psi_{N}}{\sigma_{z}^{(N)}\vert \psi_{N}} = 0$ and $\bra{\phi_{1}}\otimes \cdots \otimes \bra{\phi_{N-1}} H_{N-1} \ket{\psi_{1}}\otimes \cdots \otimes \ket{\psi_{N-1}} = \bra{\psi_{1}}\otimes \cdots \otimes \bra{\psi_{N-1}} H_{N-1} \ket{\phi_{1}}\otimes \cdots \otimes \ket{\phi_{N-1}} = 0$, it is clear that $\theta = \pi/2$ is an extremum which is easily verified to be a maximum. By the definition of $\mathcal{M}$ in Eq.\,(\ref{eqn:macrodef}), we find the value of $\mathcal{M}$ given by Eq.\,(\ref{eqn:mvalue}).
$\square$

It is clear that $\mathcal{M}$ is close to the maximal value $\vert J \vert$ if and only if $(1-\vert z_{j}\vert )^{2}(1-\vert z_{k} \vert)^{2}$ is close to 1. As discussed below, this is not the case for $z_{j} = u_{j}^{{\bm Q}_{L}}u_{j}^{{\bm Q}_{R}} +v_{j}^{{\bm Q}_{L}}v_{j}^{{\bm Q}_{R}}$, which appear when $\ket{G} = \ket{G}_{SFQ}$. This is the reason that $\mathcal{M}$ for $\ket{G}_{SFQ}$ does not achieve values of the same order of magnitude as $\vert J \vert$, where in this case, $J$ is the subset of momentum space $\Lambda$ obtained by removing a subset defined by a cutoff $k_{0}$, i.e., $J= \Lambda \setminus \lbrace \Vert \bm{k} \Vert > k_{0} \rbrace $.

In Appendix \ref{sec:upperb}, we derive the following tight upper bound for $\mathcal{M}$:
\begin{eqnarray} 
\mathcal{M} \lesssim {2\lambda \over 1+e^{-\lambda}}(1-\vert \Lambda \vert^{-1}) + 1\label{eqn:macroform} 
\end{eqnarray} 
which is a function of $\vert \langle \circlearrowleft \vert \circlearrowright \rangle \vert$ only. Although the condition $\vert \langle \circlearrowleft \vert \circlearrowright \rangle \vert \ll 1$ is satisfied by all operating bare flux qubits, the macroscopicity $\mathcal{M}$ can be further increased by manipulating an isolated or hybrid SFQ system in such a way that $\vert \langle \circlearrowleft \vert \circlearrowright \rangle \vert$ is decreased. In the remaining part of this section, we show that the relationship between $\mathcal{M}$ and $\lambda$ given by Eq.(\ref{eqn:macroform}) allows the macroscopicity to be calculated to good approximation from the action of an instanton connecting $\ket{\circlearrowleft}$, $\ket{\circlearrowright}$.

An accurate estimate of $\lambda$ in Eq.\,(\ref{eqn:macroform}) can be obtained experimentally by spectroscopic determination of spectral gap of the SFQ or theoretically from the shape of the flux potential $V(\Phi)$ with degenerate minima $\pm \Phi_{0}/2$ that occur when the externally applied flux $\Phi_{\text{ext}}$ satisfies $\Phi_{\text{ext}} = \Phi_{0}/2$, which falls in the parameter regimes explored in recent experiments demonstrating macroscopic superpositions in flux qubits \cite{lukens,wilhelm,clarke2,clarke}. To show how $\lambda$ can be estimated from the spectral gap of a SFQ, we consider a two-level, quasi-degenerate Hamiltonian $H_{\text{d}}:= 
(\gamma /2)\left( \ket{\circlearrowleft}\bra{\circlearrowleft}+ \ket{\circlearrowright}\bra{\circlearrowright} \right)$ ($\gamma >0$) which describes the dynamics of nearly orthogonal screening current states at the point $\Phi_{\text{ext}} = \Phi_{0}/2$.  The states $\ket{\circlearrowleft}$, $\ket{\circlearrowright}$ correspond to well-defined values $\Phi_{0}/2$, $-\Phi_{0}/2$, respectively, of the flux, and have the same expected energy $\langle H_{\text{d}} \rangle = (\gamma/2)(1 + e^{-2\lambda})$ with respect to $H_{\text{d}}$. The spectral gap of $H_{\text{d}}$ is found to be $\Delta E = \gamma e^{-\lambda}$. Knowledge of the expected energy $\langle H_{\text{d}} \rangle$ can be combined with the spectroscopic determination of $\Delta E$ to obtain a value for $\lambda$. On the other hand, the flux values $\pm \Phi_{0}/2$ corresponding to $\ket{\circlearrowleft}$, $\ket{\circlearrowright}$ are analogous to the degenerate minima $\pm q_{0}$ of the double-well potential that is traditionally used to illustrate instanton methods. For the present flux potential $V(\Phi)$, the spectral gap $\Delta E$ can be computed using the semiclassical method \cite{coleman}, which yields $\Delta E = 2\hbar R e^{-S_{\text{inst}}}$, where $S_{\text{inst}} = \int_{-\Phi_{0}/2}^{\Phi_{0}/2}d\Phi \sqrt{2CV(\Phi)}$ is the action of an instanton solution $\Phi$  of the imaginary time equation of motion corresponding to Eq.\,(\ref{eqn:fluxactbare}) and $R$ is a ratio of fluctuation determinants. From the above equivalent methods of computing $\Delta E$, we obtain an equation for $\lambda$:
\begin{equation}
\lambda = {S_{\text{inst}}\over \hbar} + \ln \left[\gamma \over \hbar R\right] .
\label{eqn:instantonlambda}
\end{equation}
Combined with Eq.\,(\ref{eqn:macroform}), Eq.\,(\ref{eqn:instantonlambda}) leads to a major conclusion of the present work: A decrease of the overlap between the degenerate flux states $\ket{\circlearrowleft}$, $\ket{\circlearrowright}$ (i.e., increase in $\lambda$) results in greater superposition macroscopicity $\mathcal{M}$ through the increase in the action of an instanton traveling between the degenerate potential minima in imaginary time. For the experimental demonstration of SFQ in Ref.\,\cite{lukens} in which $\Phi_{0}^{2}/2L =645 \text{ K}$, $I_{c}=(152\pi/ \Phi_{0})\text{K} $ and $E_{C}=9\times 10^{-3}\text{K}$, we find that $\mathcal{M} \lesssim 481$, which is lower than the upper bound $\mathcal{M} \le 3800$ derived by considering the displacement of Fermi surfaces of the two components of $\ket{G}_{SFQ}$ \cite{korsbakken}. For the experiment in Ref.\,\cite{wilhelm} with $(I_{c}\Phi_{0}/2\pi)/E_{C} = 38 \pm 8$ and $(\Phi_{0}^{2}/2L)/E_{C} \approx 2\times 10^{4}$, we find $\mathcal{M}\lesssim 227$.

\section{\label{sec:qcr}Metrological usefulness of flux qubit}

To relate a large value of $\mathcal{M}$ to the usefulness of $\ket{G}_{SFQ}$ as a quantum magnetometer, we note that if $H_{0}$ is the observable giving the maximum variance in Eq.\,(\ref{eqn:macrodef}), the quantum Cram\'{e}r-Rao theorem \cite{holevo} implies that a single shot unbiased estimate $\hat{\theta}$ of the phase $\theta$ in the unitary operator $\exp(-i\theta H_{0})$ satisfies 
\begin{equation}\sqrt{\langle (\Delta \hat{\theta})^{2}\rangle } \ge (4\langle (\Delta H_{0})^{2} \rangle_{\ket{\Psi}})^{-1/2} =(4 \mathcal{M}\vert J \vert)^{-1/2} .\label{eqn:cramraomacro}
\end{equation}
Increasing $\mathcal{M}$ from 1 to $\vert J \vert$ interpolates the quantum Cram\'{e}r-Rao bound between standard quantum limit precision scaling $\mathcal{O}(\vert J \vert^{-1/2})$ and Heisenberg limit precision scaling $\mathcal{O}(\vert J \vert ^{-1})$ for estimation of $\theta$.  For a measurement-imposed momentum cutoff $k_{0}$ and for small $Q$, we show in Appendix \ref{sec:metrouseappend} that the 1-local Hamiltonian which has largest variance in $\ket{G_{k_{0}}}_{SFQ}$ is approximately $H_0=\sum_{\bm k}\sigma_{z}^{({\bm k})}$, where the operator equalities $\sigma_{z}^{({\bm k})} = 1-a_{{\bm k},\uparrow}^{\dagger}a_{{\bm k},\uparrow}-a_{{\bm k},\downarrow}^{\dagger}a_{{\bm k},\downarrow}$ and $\sigma_{x}^{({\bm k})}=a_{{\bm k},\uparrow}a_{-{\bm k},\downarrow} + a_{-{\bm k},\downarrow}^{\dagger}a_{{\bm k},\uparrow}^{\dagger}$ hold in the subspace $\lbrace \ket{0}_{\bm k},\ket{1}_{\bm k} \rbrace$. The variance of the operator $\sum_{\bm k}\sigma_{z}^{({\bm k})}$ ($\sum_{\bm k}\sigma_{x}^{({\bm k})}$) corresponds to number fluctuation (order parameter fluctuation). Since the total electron number operator is the canonical conjugate of the flux operator of the SFQ, $\ket{G_{k_{0}}}_{SFQ}$ is most useful as a probe for estimation of displacements of flux through the SFQ. The quantum Cram\'{e}r-Rao bound for the error of a single-shot, unbiased estimator $\hat{\Phi}$ of the flux in $\ket{G_{k_{0}}}_{SFQ}$ is given by 
\begin{equation}
{\sqrt{\langle (\Delta \hat{\Phi})^{2} \rangle_{\ket{G_{k_{0}}}_{SFQ}}}\over \Phi_{0}} \ge {1\over \sqrt{\mathcal{M}}} {1 \over \sqrt {4  \vert \Lambda \setminus \lbrace \Vert {\bm k} \Vert > k_{0} \rbrace \vert }},
\label{eqn:qcr}
\end{equation}  
where $\vert \Lambda \setminus \lbrace \Vert {\bm k} \Vert > k_{0} \rbrace \vert$ is the number of momentum modes inside the cutoff radius.
Because $\ket{G_{k_{0}}}_{SFQ}$ exhibits a value of $\mathcal{M}$ greater than that of $\ket{\circlearrowleft}$ or $\ket{\circlearrowright}$ (each having $\mathcal{M} = 1$), its corresponding quantum Cram\'{e}r-Rao bound is reduced below the standard quantum limit by a factor of $1/\sqrt{\mathcal{M}}$. Note that Eq.\,(\ref{eqn:qcr}) implies $\mathcal{M}$ can be extracted from the measurement statistics of a flux estimator if that estimator is optimal (in the sense of achieving the quantum Cram\'{e}r-Rao bound \cite{helstrombook}); otherwise one extracts a lower bound for $\mathcal{M}$. The upper bound in Eq.\,(\ref{eqn:macroform}) can be extracted from knowledge of the parameters of flux potential $V(\Phi)$ \cite{lukens}, obtained from theoretically from the microscopic theory or experimentally from spectroscopy of the SFQ.

\section{\label{sec:effect}Effective action of inductively coupled Bose gas and flux qubit}

We now demonstrate that bringing an ultracold magnetized Bose gas into magnetic contact with the SFQ increases $\lambda$, and hence $\mathcal{M}$, in an experimentally realizable geometry (Fig.\,\ref{fig:sapphireembedding}). We thus explicitly construct a hybrid SFQ-BEC device that allows an amplified value of $\mathcal{M}$ for $\ket{G_{k_{0}}}_{SFQ}$, and thereby increases the metrological usefulness of the superconductor screening-current superposition state for quantum magnetometry.

Integration of a magnetically trapped BEC setup into a
superconducting quantum circuit has already been experimentally achieved in a similar geometry with BEC/SFQ distance $D= 17\,\mu$m \cite{weisstubingen}. The experimental realization of the amplification of $\mathcal{M}$ in the present setup requires that the SFQ be held at temperatures $T\sim 10-100$\,mK \cite{clarke}, that the Bose gas be magnetized and thermally insulated from the SFQ, and that the Bose gas be held within $D= 3\,\mu$m of the SFQ (cf. discussion after Eq.\,(\ref{eqn:kresult}) below). The fully magnetized Bose gas can be achieved by cooling a spinor Bose gas through the magnetization transition in an optical trap, and bringing it into proximity with the SFQ by, e.g., magnetic conveyor under ultrahigh vacuum,
as demonstrated in \cite{magconvey}, or by using trapping wire currents and a bias magnetic field to locally position the Bose gas. In order to exploit the amplification of $\mathcal{M}$ for magnetometry, the measurement of $\hat{\Phi}$ should be unbiased and should at least achieve the standard quantum limit for flux estimation. In addition, it must be carried out within the coherence time of $\ket{G}_{SFQ}$ ($\sim 1-10\,\mu$s \cite{Devoret,Stern}). Because of these demands, a magnetometry protocol benefitting from the amplification of $\mathcal{M}$ would likely involve a readout of the SFQ by quantum memory engineered in the Bose gas itself \cite{PhysRevLett.111.240504} or by a spin-ensemble coupled to the SFQ \cite{Saito}.

We now derive the inductance renormalization $\mathcal{K}$ due to the proximal magnetized Bose gas, which increases $\lambda$ by Eq.\,(\ref{eqn:instantonlambda}) and thereby increases $\mathcal{M}$ of the state $\ket{G}_{\text{SFQ}}$ by Eq.\,(\ref{eqn:macroform}). 
In the vicinity of a superconducting tunnel junction, the SFQ consists of a left region $V_{S}^{(L)}$ and a right region $V_{S}^{(R)}$ of $s$-wave, type-II superconductor with characteristic radius $R_{S}$. The spinor Bose gas is trapped in a volume $V_{B}$ which we assume can be brought to within a few multiples of $R_{S}$ from the plane of the SFQ (see Fig.\,(\ref{fig:sapphireembedding})).  To calculate the effective action of the flux instanton in this geometry, we use the bosonic/fermionic imaginary time coherent state path integral describing the dynamics of the electron (Nambu) fields $\Psi= (\psi_{\uparrow},\psi_{\downarrow}^{*})^{T}$, $\Psi^{*}=(\psi_{\downarrow}^{*},\psi_{\uparrow})$ and the complex bosonic fields ${\bm \phi}$, $\overline{\bm \phi}$ corresponding to a spin-$F$ atomic Bose gas. Integrating over the electron fields leaves only the flux and the atoms as quantum degrees of freedom. Building upon the microscopic derivation of the capacitive, inductive, and sinusoidal terms of the effective flux action \cite{ambegaokar}, we focus on the effective contribution arising from the Bose gas--SFQ interaction, which is taken to be a linear Zeeman coupling of the magnetic moment of the spinor Bose gas and the magnetic field produced by the supercurrent in the SFQ
\cite{PhysRevA.87.052303,PhysRevLett.111.240504}.

The magnetic field in the Bose gas due to the superconducting currents is calculated from the Biot-Savart law using the gauge-invariant superconducting current $
{\bm J} = (e \hbar / 2m_{e}) \sum_{\sigma \in \pm 1/2}\psi_{\sigma}^{*}\left( -i\nabla - {e\over \hbar } {\bm A} \right) \psi_{\sigma} + {\rm G.c.}$, where $\rm G.c.$ signifies the Grassmann conjugate of the preceding term.
The magnetic interaction can be simplified by implementing the local change of basis $\psi_{\sigma} \mapsto \psi_{\sigma}e^{-i\theta_{L/R}/2}$, $\psi_{\sigma}^{*} \mapsto \psi_{\sigma}e^{i\theta_{L/R}/2}$ on the electron fields, where $\theta_{L/R}$ is the phase field of the superconductor order parameter in the left or right regions of the SFQ proximal to the tunnel junction. This rotation allows us to consider a spatially constant, temperature-dependent value for the modulus of the superconductor order parameter $\Delta_{L}=\Delta_{R}=\Delta$ obtained from the BCS gap equation while retaining the quantum fluctuations of the order parameter phase \cite{abrikosov}.  In addition, we introduce the gauge-invariant velocity ${\bm v}_S:= - \hbar /2m_{e}(\nabla \theta + {2e \over \hbar }{\bm A})$, where $\theta$ smoothly varies between the values $\theta_{L}$ and $\theta_{R}$ achieved at the respective left and right of the tunnel junction, which is connected to the flux degree of freedom via 
\begin{equation}
\left\Vert \int_{C} d\ell \cdot {\bm v}_S \right\Vert = {e \over m_{e}} \Phi = {\pi \hbar \over m_{e}}{\Phi \over \Phi_{0}},
\label{eqn:fluxoid}
\end{equation}
according to the fluxoid quantization condition \cite{degennes}.

The full form of the interaction in Nambu space is
\begin{eqnarray}
S_{\text{int}}&=&-C_{1}\sum_{m=L}^{R} \int_{0}^{\beta \hbar }\! dx_{0}\, \int_{V_{B}}^{({\bm x})}\, \overline{\bm \phi}\, \mathuu{\sigma_{i}} {\bm \phi} \, \epsilon^{ij \ell} \int_{V_{S}^{(m)}}^{({\bm x}')} \!   \del_{\ell}\Vert {\bm x}-{\bm x}' \Vert^{-1}  \nonumber \\ &{}& 
\times \Psi^{*}_{m} \left[ \left( i\overleftarrow{\del_{j}'} - i\del_{j}' \right)\mathbb{I} + {2m_{e} \over \hbar }v_{S,j} \tau_{z} \right] \Psi_{m},
\label{eqn:microint}
\end{eqnarray}
where $C_{1}=g_{f}\mu_{B}\mu_{0}e \hbar / 8\pi m_{e}$ in SI units, $\mathuu{{\bm \sigma} } = (\sigma_{x},\sigma_{y}, \sigma_{z})$ is a vector of spin-$F$ operators, the spatial integrals over $V_{B}$ and $V_{S}^{(m)}$ are labeled with their respective coordinates ${\bm x}$ and ${\bm x}'$, and summation over repeated indices is implied. The Nambu space Pauli operators $\mathbb{I}$, $\tau_{x}$, $\tau_{y}$, $\tau_{z}$ have been introduced. We leave the dynamics of the spinor Bose gas unspecified, as we eventually only require a fully magnetized state, which exists as an eigenstate of several atomic Hamiltonians.

In the self-consistent BCS mean field theory, the superconducting ($S_{SC}$) and interaction ($S_{\text{int}}$) contributions to $S_{\text{tot}}$ are quadratic in the Nambu spinors. Upon performing the Gaussian integral over $\Psi$, $\Psi^{*}$, the effective action becomes  $S_{\text{eff}}^{(2)} =  
-\frac12\text{tr}\left[{G_{0}}\left(G_{\text{int}}^{-1}+\sum_{j}G^{-1}_{j}\right){G_{0}}\left(G_{\text{int}}^{-1}+\sum_{j}G^{-1}_{j}\right)\right] $ at lowest nonzero order of expansion of the functional determinant in terms of $G_{\text{int}}^{-1}$ (Appendix \ref{sec:effectderivation}). Here, $G_{\text{int}}^{-1}$ is the integral kernel arising from Eq.\,(\ref{eqn:microint}) and $G_{j}^{-1}$ are the kernels arising from the chemical potential across the tunnel junction, the kinetic energy of the superfluid velocty ${\bm v}_S$, and the tunneling amplitude, respectively. $G_{0}$ is the free 2$\times$2 Gor'kov Green function of the superconductor and the trace is taken over all internal momenta and Matsubara frequency in addition to the Nambu space indices. The first order contribution of $G_{\text{int}}^{-1}$ vanishes, along with the second order terms of the form $\text{tr}\left[G_{0}G_{\text{int}}^{-1}G_{0}G_{j}^{-1}\right]$ (see Appendix \ref{sec:effectderivation}). 
The tunneling contribution to $S_{\text{eff}}^{(2)}$ is nonlocal in imaginary time, giving rise to dissipative real time evolution of the flux degree of freedom. We will show that time nonlocality also appears in the interaction contribution $S_{\text{eff,int}}^{(2)}$ to $S_{\text{eff}}^{(2)}$ and then proceed to study the low-temperature (coherent) limit.

The first term in brackets in Eq.\,(\ref{eqn:microint}) vanishes because only terms that are even in momentum contribute to the sum over internal momenta in the expression $S_{\text{eff,int}}^{(2)}=-\frac12\text{tr}({G_{0}(\omega_{n},{\bm k})}G_{\text{int}}^{-1}({\bm k},\omega_{n};{\bm k}',\omega_{m})$ ${G_{0}({\bm k}',\omega_{m})}G_{\text{int}}^{-1}({\bm k}',\omega_{m};{\bm k},\omega_{n}))$. We make the approximation ${\bm k}={\bm k}'$ which amounts to neglecting scattering of Bogoliubov quasiparticles from the superconducting superfluid current.
This approximation results in a phase space factor proportional to the energy scale of the attractive BCS electron pairing: $\int d^{3}k'/(2\pi)^{3} \approx (\rho(E_{F})/4\pi^{2}) \int_{-\hbar \omega_{D}}^{\hbar \omega_{D}}d\epsilon$, where we have used $\epsilon_{\bm k} = {\hbar^{2}\Vert {\bm k}\Vert^{2}\over 2m_{e}} - \mu$ as the free electron dispersion ($\mu = \mu_{L} = \mu_{R} \approx E_{F}$ is the $T=0$ chemical potential of the SFQ), $\omega_{D}$ is the Debye frequency 
and $\rho(E_{F})$ is the density of quasiparticle states at the Fermi surface. The intermediate expression for the effective action due to $G_{\text{int}}^{-1}$ is, with implicit integrals over $ x_{0}$, $x_{0}'$ assumed, given by 
\begin{eqnarray}
&{}&-{C_{2}\over (\beta \hbar)^{2}}\, \int^{({\bm y})}_{V_{B}} \int {d^{3}q\over (2\pi)^{3}} {{\bm M}(x_{0},{\bm y}) \cdot ({\bm v}_S (x_{0},{\bm q}) \times {\bm q}) e^{i{\bm q}\cdot {\bm y}}\over \Vert {\bm q}\Vert^{2}} \nonumber \\ &{}& \int^{({\bm y'})}_{V_{B}} \int {d^{3}q'\over (2\pi)^{3}} {{\bm M}(x_{0}',{\bm y}') \cdot ({\bm v}_S(x_{0}',{\bm q}') \times {\bm q}') e^{i{\bm q'}\cdot {\bm y'}}\over \Vert {\bm q'}\Vert^{2}} \nonumber \\ &{}& \int_{-\hbar \omega_{D}}^{\hbar \omega_{D}}d\epsilon \sum_{(n,r)\in \mathbb{Z}\times \mathbb{Z}}S(\omega_{n},\omega_{r},\epsilon)e^{i\omega_{n}(x_{0}'-x_{0})}e^{i\omega_{r}(x_{0}-x_{0}')} ,
\label{eqn:intermed}
\end{eqnarray}
where ${\bm M}(x_{0},{\bm y}):= \overline{{\bm \phi}}(x_{0},{\bm y})\mathuu{\bm{ \sigma}} {\bm \phi}(x_{0},{\bm y})$, $S(\omega_{n},\omega_{r},\epsilon)$ is the temperature-dependent factor arising from the matrix trace of the free Green's functions, and \begin{equation}
C_{2} = (g_{f}\mu_{B}\mu_{0}e)^{2}\rho(E_{F})^{2}\hbar \omega_{D} / 8\pi^{4}.\label{eqn:c2}\end{equation}  
The integrand of Eq.\,(\ref{eqn:intermed}) is nonlocal in time and can be computed in terms of the Bickley function $\text{Ki}_{1}(0)$ \cite{bickleynayler}. For the present purpose of calculating the inductance renormalization $\mathcal{K}$ in Eq.\,(\ref{eqn:fluxactbare}), we evaluate the Matsubara sum and take the coherent part of Eq.\,(\ref{eqn:intermed}), i.e., $x_{0}=x_{0}'$. After an inverse Fourier transform, we assume that $({\bm x}-{\bm x}')\cdot (\langle {\bm M}(x_{0},{\bm x}) \rangle_{\ket{\text{BEC}}} \times {\bm v}_S(x_{0},{\bm x}') )\approx D \Vert \langle {\bm M}(x_{0},{\bm x}) \rangle_{\ket{\text{BEC}}} \Vert \Vert {\bm v}_S(x_{0},{\bm x}') \Vert $ in the geometry of Fig.\,\ref{fig:sapphireembedding}, where $\ket{\text{BEC}}$ represents a condensed state of the Bose gas such that $\int_{V_{B}} d^{3}x\Vert \langle {\bm M}(x_{0},{\bm x}) \rangle_{\ket{\text{BEC}}} \Vert \in \mathcal{O}(N_{B})$ with $N_{B}$ the total number of atoms. Making use of Eq.\,(\ref{eqn:fluxoid}) by approximating $\int_{V_{S}^{(L)}\sqcup V_{S}^{(R)}}d^{3}x\Vert {\bm v}_S(x_{0},{\bm x}') \Vert \approx (\pi \hbar R_{S}^{2}/ m_{e}) \Phi(x_{0})/\Phi_{0}$, the effective action is
\begin{equation}
S_{\text{eff,int}}^{(2)} = {\pi \hbar C_{2} \over 2^{5}}\left({R_{S} \over D }\right)^{4}N_{B}^{2}\int_{0}^{\beta \hbar} dx_{0}{\Phi(x_{0})^{2} \over \Phi_{0}^{2}}.
\label{eqn:secondordergint}
\end{equation}
We then have in Eq.\,(\ref{eqn:fluxactbare})
\begin{equation}
\mathcal{K} = {\pi \hbar C_{2}\over 2^{5}\Phi_{0}^{2}}\left({R_{S} \over D }\right)^{4}N_{B}^{2}.
\label{eqn:kresult}
\end{equation}
To estimate feasible values for $\mathcal K$, 
we use $\rho(E_{F}) = 4.58 \times 10^{46} \text{J}^{-1}\text{m}^{-3}$, $|g_{f}|=2$, and $\hbar \omega_{D} = 3.21 \times 10^{-20}\text{J}$ for Al, which results in $\pi \hbar C_{2} /2^{5}= 1.57 \times 10^{-31}\text{J}$. If $R_{S}=1\, \mu\text{m}$ and $D=3\, \mu\text{m}$, which are within reach for hybrid systems composed of SFQ and Bose gas \cite{weisstubingen}, then $N_{B}=2\times 10^{6}$ condensed atoms are sufficient for the renormalized inductive energy $E_{L'}:=\Phi_{0}^{2}\mathcal{K}$ to approximately equal the bare inductive energy $E_{L}$ of the SFQ in Ref.\,\cite{lukens}. Using Eq.\,(\ref{eqn:instantonlambda}) and $E_{L} + \Phi_{0}^{2}\mathcal{K}= 2\times 645\,$K now gives $\mathcal{M}\lesssim 677$. In the more extreme case of $R_{S}=D$, $N_{B} = 5 \times 10^{6}$, $\mathcal{M}\lesssim 3114$ which gives a lower quantum Cram\'{e}r-Rao bound for external flux estimation by a factor of $1/2$ compared to $R_{S} / D = 1/3$. 
The renormalized action for an instanton, and hence the macroscopicity $\mathcal{M}$ [see Eq.\,(\ref{eqn:macroform})], scales as $\mathcal{O}(N_{B})$. 
From this analysis, we see that a screening current superposition $\ket{G}_{SFQ}$ in the SFQ of the hybrid system has a twofold to fivefold larger value of $\mathcal{M}$ than the analogous state of the non-hybrid SFQ. Assuming an optimal measurement on the superconductor can be carried out, the hybrid system can therefore act as a quantum magnetometer operating further below the standard quantum limit than is possible for probe states of the SFQ in the absence of the magnetized Bose gas.

We note that the inductance renormalization has important consequences for the operation of the SFQ. In order to exhibit the nonlinear potential defining the qubit, the inductance of the Josephson junction $\Phi_{0}( I_{c}\cos \Delta \gamma)^{-1}$ should cancel the renormalized inductance of the loop \cite{leshouchesdevoret}. This requires that the expected phase difference $\Delta \gamma$ across the junction must increase toward $\pi /2$. A large value of $\mathcal{M}$ can thus be interpreted physically as a result of the large macroscopic phase difference across the tunnel junction maintained in the $\ket{\circlearrowleft}$, $\ket{\circlearrowright}$ flux branches. 

\section{Conclusions}

In summary, we have shown that the macroscopicity $\mathcal{M}$ of the SFQ ground state $\ket{G}_{SFQ}$ is amplified by coupling to a proximal magnetized spinor Bose gas. The required magnetization can be achieved by preparation of the spinor Bose gas in a polar Bose-Einstein condensate phase. It is useful to note that the lack of electrical conductivity of the Bose gas and the tunability of the gas magnetization by control of the number of trapped atoms and control of the transition into the polar phase  make the the SFQ-BEC hybrid system considered in the present work a superior setting for hybrid quantum magnetometry compared to, e.g., a hybrid system of a SFQ and a static metallic magnet.

Amplification of $\mathcal{M}$ in the hybrid system indicates the greater theoretical precision obtainable (compared to the ground state superposition of a non-hybrid SFQ) when $\ket{G}_{SFQ}$ is used to probe magnetic displacements. We expect this result to stimulate further research on the quantum macroscopicity of hybrid persistent current qubits and on optimal preparation and measurement protocols for
achieving the quantum Cram\'{e}r-Rao bound for magnetometry in hybrid systems. 

\begin{acknowledgements}
We thank J.\,\,Fort\'agh for a helpful discussion on the experimental realization of our theoretical proposal. 
This research was supported by the NRF Korea, Grant No. 2014R1A2A2A01006535.
\end{acknowledgements}

\appendix

\section{\label{sec:upperb}Proof of upper bound for $\mathcal{M}$ in Eq.\,(\ref{eqn:macroform})}

Consider $\lambda = -\log \vert \langle \circlearrowleft \vert \circlearrowright \rangle \vert = -\log \left( \prod_{\bm k}u_{\bm k}^{{\bm Q}_{L}}u_{\bm k}^{{\bm Q}_{R}} +v_{\bm k}^{{\bm Q}_{L}}v_{\bm k}^{{\bm Q}_{R}} \right) $ in the simple case of a superposition of uniform supercurrent, e.g., with ${\bm Q}_{R}=0$ and variable $\bm{Q}_{L}$ satisfying $\Vert {\bm Q}_{L} \Vert \ll \sqrt{2m_{e}\Delta}/ \hbar$. By calculating $u_{\bm k}^{{\bm Q}_{L}}$, $u_{\bm k}^{0}$, $v_{\bm k}^{{\bm Q}_{L}}$, and $v_{\bm k}^{0}  $ in the vicinity of the Fermi momentum $k= k_{F}$ (see Eq.\,(\ref{eqn:coeffbcs})), one finds that $u_{\bm k}^{{\bm Q}_{L}}u_{\bm k}^{0} +v_{\bm k}^{{\bm Q}_{L}}v_{\bm k}^{0} \approx 1$ \cite{degennes}. Introducing the real numbers $x_{\bm k}$ satisfying $0<x_{\bm k}\ll 1$ defined by $x_{\bm k}:= 1-(u_{\bm k}^{{\bm Q}_{L}}u_{\bm k}^{0} +v_{\bm k}^{{\bm Q}_{L}}v_{\bm k}^{0})^{2} $ for each $\bm{k}$, it follows from the series expansion of the logarithm that \begin{equation} 2\lambda = \sum_{\bm k}x_{\bm k} + {1\over 2}\sum_{\bm k}x_{\bm k}^{2}+\ldots . \end{equation} As a consequence of Proposition 1 in Sec. \ref{sec:macrosec}, the macroscopicity $\mathcal{M}$ of $\ket{G}_{SFQ}$ is $\mathcal{M} = 1+ \sum_{{\bm k}\neq {\bm k'}}\sqrt{x_{\bm k}x_{{\bm k'}}} /\vert \Lambda \vert (1+e^{-\lambda})$, 
where $\vert \Lambda \vert$ is the volume of momentum space for which Cooper pairing occurs. By the arithmetic mean-geometric mean inequality and the fact that $  \sum_{\bm k}x_{\bm k} \le 2\lambda $, it follows that $  \sum_{{\bm k}\neq {\bm k'}}\sqrt{x_{\bm k}x_{\bm k'}} \le \sum_{{\bm k}\neq {\bm k'}} {x_{\bm{k}}+x_{\bm{k'}} \over 2}  \le 2(\vert \Lambda \vert -1)\lambda$ and therefore,
\begin{eqnarray} 
\mathcal{M} \le {2\lambda \over 1+e^{-\lambda}}(1-\vert \Lambda \vert^{-1}) + 1.
\label{eqn:a2eqn}
\end{eqnarray} 
Assuming that $\Vert {\bm Q}_{L} \Vert \ll \sqrt{2m_{e}\Delta}/ \hbar$, the quantity $u_{\bm{k}}^{\bm{Q}_{L} }u_{\bm{k}}+ v_{\bm{k}}^{\bm{Q}_{L}}v_{\bm{k}}$ does not vary significantly over $\bm{k}$  near the Fermi surface and is nearly equal to 1. As a consequence, $\sum_{{\bm k}\neq {\bm k'}}\sqrt{x_{\bm k}x_{\bm k'}}$ is well-approximated by $(\vert \Lambda \vert - 1 ) \sum_{{\bm k}} x_{\bm k} = (\vert \Lambda \vert - 1 )\left( 2\lambda - \mathcal{O}(\sum_{\bm{k}}x_{\bm{k}}^{2})\right)$ and the bound in Eq.(\ref{eqn:a2eqn}) is tight. Quantitatively, the condition $\vert x_{\bm{k}}-x_{\bm{k'}}\vert < \vert \Lambda \vert^{-1}$ for all $\bm{k},\bm{k'} \in \Lambda$ is sufficient to guarantee that the difference between $\mathcal{M}$ and the upper bound of Eq.(\ref{eqn:a2eqn}) is less than 1. Note that $\lambda$ can increase if and only if $\sum_{{\bm k}\neq {\bm k'}}\sqrt{x_{\bm k}x_{\bm k'}}$ increases. Therefore, as the superposed BCS states of $\ket{G}_{SFQ}$ become orthogonal, the macroscopicity $\mathcal{M}$ increases.

\section{\label{sec:metrouseappend}Metrological usefulness of $\ket{G}_{SFQ}$}

Here we determine the 1-local, norm one operator $H$ which exhibits maximal variance in $\ket{G}_{SFQ}$ (we take ${\bm Q'}=0$, and ${\bm Q} \neq 0$ for simplicity). According to the proof of Proposition 1 
\begin{widetext}
\begin{eqnarray}
H&=&\sum_{{\bm k} \in \Lambda} e^{-i\tan^{-1}(v_{\bm k}/u_{\bm k})\sigma_{y}^{({\bm k})}}\ket{0}_{\bm k}\bra{0}_{\bm k}e^{i\tan^{-1}(v_{\bm k}/u_{\bm k})\sigma_{y}^{({\bm k})}} - e^{-i\tan^{-1}(v_{\bm k}^{{\bm Q}}/u_{\bm k}^{{\bm Q}})\sigma_{y}^{({\bm k})}}\ket{0}_{\bm k}\bra{0}_{\bm k}e^{i\tan^{-1}(v_{\bm k}^{{\bm Q}}/u_{\bm k}^{{\bm Q}})\sigma_{y}^{({\bm k})}} \nonumber \\
&=& \sum_{{\bm k}\in \Lambda} c_{z}^{({\bm k})}\sigma_{z}^{({\bm k})} + c_{x}^{({\bm k})}\sigma_{x}^{({\bm k})}
\end{eqnarray} where $\left( c_{z}^{({\bm k})} \right)^{2} + \left( c_{x}^{({\bm k})} \right)^{2} = 1$ for all ${\bm k}$ and
\begin{eqnarray}
c_{z}^{({\bm k})} &:=& {\left(  u_{\bm k}^{2}- \left(  u_{\bm k}^{{\bm Q}}\right)^{2}\right) \over  \sqrt{1- ( u_{\bm k}u_{\bm k}^{{\bm Q}} - v_{\bm k}v_{\bm k}^{{\bm Q}} )^{2}} }\nonumber \\ c_{x}^{({\bm k})} &:=& {(u_{\bm k}v_{\bm k} - u_{\bm k}^{{\bm Q}}v_{\bm k}^{{\bm Q}}) \over \sqrt{1-(u_{\bm k}u_{\bm k}^{{\bm Q}} - v_{\bm k}v_{\bm k}^{{\bm Q}})^{2}} }.
\label{eqn:ccoeff}
\end{eqnarray}
The coefficient $c_{z}^{({\bm k})}$ is plotted in Fig.\,(\ref{fig:maxvarop}).
\begin{center}
\begin{figure}
\includegraphics[scale=0.45]{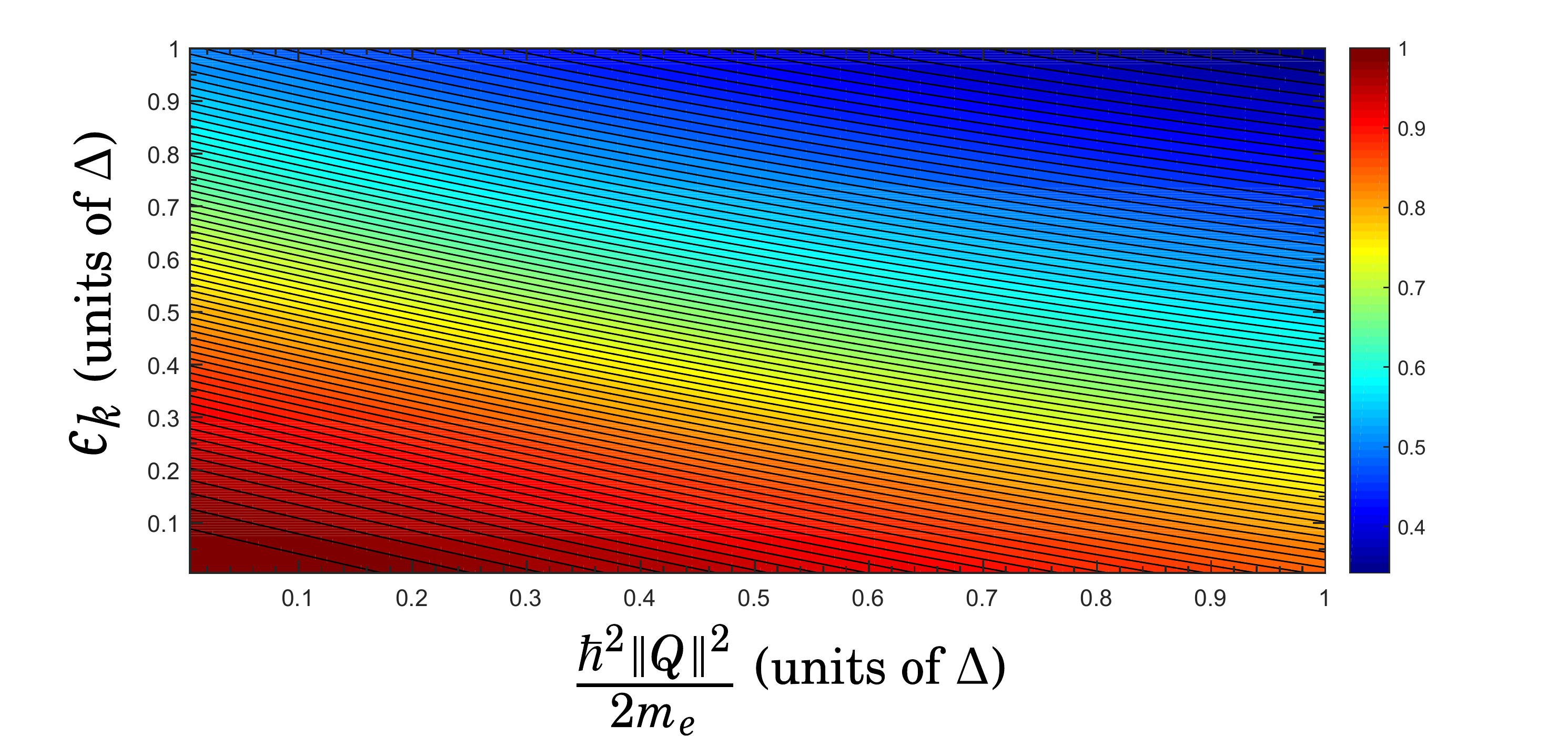}{\centering}
\caption{The coefficient $c_{z}^{({\bm k})}$ appearing in the 1-local, unit norm Hamiltonian $H$ maximizing $\langle H^{2} \rangle_{\ket{G}} - \langle H \rangle_{\ket{G}}^{2}$ in terms of $\epsilon_{\bm k}:= {\hbar^{2}\Vert {\bm k}\Vert^{2} / 2m_{e}} - E_{F}$ and $\hbar^{2}\Vert {\bm Q}\Vert ^{2} / 2m_{e}$, where the latter energies are given in units of the superconductor gap $\Delta$. \label{fig:maxvarop}}
\end{figure}
\end{center}
The explicit expressions for the amplitudes appearing in the variational BCS ground states with center of mass momentum pairing at ${\bm Q}=0$ and ${\bm Q} \neq 0$, respectively, are
\begin{eqnarray}
v_{\bm k}^{{\bm Q}} &:=& {\Delta \over \sqrt{ \Delta^{2} + \left( {\hbar^{2}\Vert {\bm k} \Vert^{2}\over 2m_{e}} + {\hbar^{2}\Vert {\bm Q} \Vert^{2}\over 2m_{e}} - E_{F} + \sqrt{ \left( {\hbar^{2}\Vert {\bm k} \Vert^{2}\over 2m_{e}} + {\hbar^{2}\Vert {\bm Q} \Vert^{2}\over 2m_{e}} - E_{F} \right)^{2} + \Delta^{2}}  \right)^{2}  } } \nonumber \\
u_{\bm k}^{{\bm Q}} &:=& { {\hbar^{2}\Vert {\bm k} \Vert^{2}\over 2m_{e}} + {\hbar^{2}\Vert {\bm Q} \Vert^{2}\over 2m_{e}} - E_{F} + \sqrt{ \left( {\hbar^{2}\Vert {\bm k} \Vert^{2}\over 2m_{e}} + {\hbar^{2}\Vert {\bm Q} \Vert^{2}\over 2m_{e}} - E_{F} \right)^{2} + \Delta^{2}} \over \sqrt{ \Delta^{2} + \left( {\hbar^{2}\Vert {\bm k} \Vert^{2}\over 2m_{e}} + {\hbar^{2}\Vert {\bm Q} \Vert^{2}\over 2m_{e}} - E_{F} + \sqrt{ \left( {\hbar^{2}\Vert {\bm k} \Vert^{2}\over 2m_{e}} + {\hbar^{2}\Vert {\bm Q} \Vert^{2}\over 2m_{e}} - E_{F} \right)^{2} + \Delta^{2}}  \right)^{2}  } }.
\label{eqn:coeffbcs}
\end{eqnarray} We have $u_{\bm k}=u_{\bm k}^{{\bm Q}=0}$, $v_{\bm k}=v_{\bm k}^{{\bm Q}=0}$ and $u_{\bm k}^{2}+v_{\bm k}^{2} = \left( u_{\bm k}^{\bm Q}\right)^{2} + \left(v_{\bm k}^{\bm Q}\right)^{2} = 1$ for all ${\bm k}$ as required by normalization of each single Cooper pair state. The value of $c_{z}^{({\bm k})}$ from Eq.\,(\ref{eqn:ccoeff}) is plotted in Fig.\,\ref{fig:maxvarop}. Consider the superposition $\ket{G_{k_{0}}}_{SFQ}$ cut off at momentum $k_{0}$ such that $c_{z}^{({\bm k})} \approx 1$ for $\Vert \bm{k}\Vert \le k_{0}$: 
\begin{equation}
\ket{G_{k_{0}}}_{SFQ} := {1\over \sqrt{\mathcal{N}}} \left[ \bigotimes_{\Vert {\bm k} \Vert < k_{0}} \left( u_{\bm k}\ket{0}_{\bm k} + v_{\bm k}\ket{1}_{\bm k} \right) + \bigotimes_{\Vert {\bm k} \Vert < k_{0}} ( u_{\bm k}^{{\bm Q}}\ket{0}_{\bm k} + v_{\bm k}^{{\bm Q}}\ket{1}_{\bm k} ) \right].
\end{equation}
\end{widetext}
Then for $\Vert{\bm Q}\Vert \ll \sqrt{2m_{e}\Delta}/\hbar$ and $k_{0} \ll \sqrt{2m_{e}\Delta}/\hbar$, the observable $\sum_{\Vert {\bm k} \Vert < k_{0}}\sigma_{z}^{({\bm k})}$ gives the largest variance in the superposition $\ket{G_{k_{0}}}_{SFQ}$ over all 1-local, unit norm Hamiltonians. Flux displacement is generated by the canonically conjugate operator $\sum_{ {\bm k} }\sigma_{z}^{({\bm k})}$. Most of the terms in $\sum_{ {\bm k} }\sigma_{z}^{({\bm k})}$ correspond to $\Vert{\bm k}\Vert$ in a small neighborhood $[k_{F}-k_{0},k_{F}+k_{0}]$ due to the large density of states at the Fermi surface.  Hence, the probe state $\ket{G_{k_{0}}}_{SFQ}$ is more useful for estimation of flux displacements than any other parameter displacements imprinted on the state by unitary evolution generated by a 1-local, unit norm operator. The quantum Cram\'{e}r-Rao bound for a single-shot, unbiased measurement $\hat{\Phi}$ of the flux is then given by the result in Eq.\,(\ref{eqn:qcr}) where $\vert \Lambda \setminus \lbrace \Vert {\bm k} \Vert > k_{0} \rbrace \vert$ is the number of momentum modes inside the cutoff radius.

\section{\label{sec:effectderivation}Derivation of first order and second order contributions of $G_{\text{int}}^{-1}$}

We consider the total partition function $Z = \int \mathcal{D}[\cdots] \exp \left[ -{1\over \hbar}\int_{0}^{\beta \hbar}S_{\text{tot}} \right] $, where $S_{\text{tot}}$ is the total action and $ \int \mathcal{D}[\cdots]$ symbolizes the path integral over bosonic (complex) fields ${\bm \phi}$, $\overline{{\bm \phi}}$ with periodic boundary condition on $[0,\beta \hbar]$ and over all fermionic (Grassmann) fields $\Psi= (\psi_{\uparrow},\psi_{\downarrow}^{*})^{T}$, $\Psi^{*}=(\psi_{\downarrow}^{*},\psi_{\uparrow})$ with antiperiodic boundary condition on $[0,\beta \hbar]$.
\begin{widetext}
\begin{eqnarray}
S_{SC} &:=&\int_{0}^{\beta \hbar}dx_{0} \sum_{m=L,R}\left[ \int_{V_{S,m}}^{(\bm x_{m})} \sum_{\sigma ,\sigma ' \in \pm 1/2}\left( \psi^{*}_{\sigma}(\hbar \del_{0} -{i\over 2}\del_{0}\theta + ie\varphi - \mu )\psi_{\sigma}  + {\hbar^{2}\over 2m_{e}}(i\nabla - {e \over \hbar}{\bm A} )\psi^{*}_{\sigma}(-i\nabla - {e\over \hbar } {\bm A})\psi_{\sigma}\right) \right. \nonumber \\ &-& \left.  \left( \vert \Delta \vert \psi^{*}_{\uparrow}\psi^{*}_{\downarrow} + \vert \Delta \vert \psi_{\uparrow}\psi_{\downarrow} \right) + {1\over g}\vert \Delta \vert^{2} \right] +\int_{V_{S,L}}^{(\bm x_{L})}\int_{V_{S,R}}^{(\bm x_{R})} \sum_{\sigma \in \pm 1/2} \psi^{*}_{\sigma}({\bm x_{L}})T_{{\bm x_{L}},{\bm x_{R}}}e^{i\Delta \gamma \over 2}\psi_{\sigma}({\bm x_{R}}) + g.c.
\end{eqnarray}
where all fields are defined at the same imaginary time. We do not specify the dynamics of the spinor Bose gas. The Grassmann integral is performed to second order by using   \begin{eqnarray} \int \mathcal{D}[\Psi^{*},\Psi]\exp \left[ -{1\over \hbar }\int_{0}^{\beta \hbar} \left( S_{\text{SC}} + S_{\text{int}} \right) \right] &=:& \int \mathcal{D}[\Psi^{*},\Psi]\exp \Psi^{*}\left( G_{0}^{-1}+G_{\dot{\theta}}^{-1}+G_{T}^{-1} + G_{v_{S}}^{-1} + G_{\text{int}}^{-1} \right)\Psi \nonumber \\ &\propto& \exp \left[ \text{tr} \left( G_{0}(\sum_{j}G_{j}^{-1} ) - {1\over 2}\text{tr}G_{0}(\sum_{j}G_{j}^{-1})G_{0}(\sum_{j}G_{j}^{-1}) + \mathcal{O}(\sum_{j}G_{j}^{-1})^{3}) \right) \right] \nonumber \end{eqnarray} as an exponential of a quadratic form in Nambu space, where the $G_{j}^{-1}$ are given in Nambu space in terms of momenta and Matsubara frequency as follows:
\begin{eqnarray}
G_{\dot{\theta}}^{-1}(\omega_{n},{\bm k};\omega_{m},{\bm k}') &=& {1\over(2\pi)^{6}(\beta \hbar)^{2}}\left( {-\hbar (\omega_{n}-\omega_{m} )\over 2}\theta(\omega_{n}-\omega_{m},{\bm k} - {\bm k}') - ie\varphi(\omega_{n}-\omega_{m},{\bm k} - {\bm k}')\right)\tau_{z} \nonumber \\
G_{{\bm{v}_{S}}}^{-1}(\omega_{n},{\bm k};\omega_{m},{\bm k}') &=& {1\over(2\pi)^{6}(\beta \hbar)^{2}}  \left(-\hbar {\bm v}_{S}(\omega_{n}-\omega_{m},{\bm k}  - {\bm k}')\cdot {\bm k}'  \right)\mathbb{I}  \nonumber \\ &{}& - {1\over(2\pi)^{6}(\beta \hbar)^{2}} \left({m \over 2(2\pi)^{3}\beta \hbar}\sum_{r \in \mathbb{Z}}\int d^{3}q \, {\bm v}_{S}(\omega_{r},q)\cdot {\bm v}_{S}(\omega_{n}-\omega_{m}-\omega_{r},{\bm k} - {\bm k}' - {\bm q}) \right)\tau_{z}  .
\label{eqn:perturbs}
\end{eqnarray}
$G_{\text{int}}^{-1}$ is given by Eq.\,(\ref{eqn:microint}) in the main text. In momentum space, it is given by
\begin{eqnarray}
G_{\text{int}}^{-1}(k,\omega_{n};k',\omega_{m}) &=&  -i\left( {g_{f}\mu_{B}\mu_{0}e \hbar \over 2 m_{e}}\right) \int_{0}^{\beta \hbar } dx_{0}\left( \int_{V_{B}}d^{3}x \overline{\bm \phi} \mathuu{\sigma_{i}} {\bm \phi} \right)\epsilon^{ij \ell}   \nonumber \\ &{}&  \left[ {1 \over (2\pi)^{6}(\beta \hbar)^{2}}{e^{ix_{0}(\omega_{m}-\omega_{n})}e^{-i({\bm k} -{\bm k'})x} \over \Vert {\bm k} - {\bm k'} \Vert^{2}}(k_{\ell}-k_{\ell}')(k_{j}+k_{j}')\mathbb{I} \right. \nonumber \\ &+& \left. {2m_{e} \over \hbar}{1 \over (2\pi)^{9}(\beta \hbar)^{3}}\int d^{3}q \sum_{s\in \mathbb{Z}}{e^{ix_{0}(\omega_{s}-(\omega_{n}-\omega_{m}))}e^{-i({\bm k}-{\bm k'}-{\bm q})\cdot {\bm x}} \over \Vert {\bm k} - {\bm k'} -{\bm q} \Vert^{2}}(k_{\ell}'-k_{\ell}'' - q_{\ell})v_{S,j}(\omega_{s},{\bm q})\tau_{z} \right].
\end{eqnarray}

We omit writing out the expression for $G_{T}^{-1}$ since the second order cross-term $-{1\over 2}\text{tr}(G_{0} G_{\text{int}}^{-1}G_{0}G_{T}^{-1})$ can be shown to vanish by a simple argument, viz., $G_{\text{int}}^{-1}$ is nonzero if and only if its momentum arguments are on the same side of the Josephson junction while $G_{T}^{-1}$ is nonzero if and only if its momentum arguments are on different sides of the Josephson junction.

$G_{0}$ is local in momentum space and is given by the well-known expression
\begin{equation}
G_{0}(\omega_{n},{\bm k};\omega_{m} , {\bm k'})= {(2\pi)^{3}(\hbar \beta) \over \hbar^{2}\omega_{n}^{2} + \epsilon_{{\bm k}}^{2} +\Delta^{2}}\left( i\hbar \omega_{n} \mathbb{I} -\epsilon_{\bm k}\tau_{z} + \Delta \tau_{x}\right) \delta({\bm k}-{\bm k'})\delta_{n,m}
\end{equation}
where $\epsilon_{\bm k} := {\hbar^{2}\Vert {\bm k} \Vert^{2} / 2m_{e}} - E_{F}$. The matrix trace of $G_{0}(\omega_{n},{\bm k};\omega_{m} , {\bm k'})$ is $ {(2\pi)^{3}(\hbar \beta)(2i\hbar \omega_{n}) \over \hbar^{2}\omega_{n}^{2} + \epsilon_{{\bm k}}^{2} +\Delta^{2}}\delta_{n,m}\delta({\bm k}-{\bm k'})$ while the  matrix trace of $G_{0}(\omega_{n},{\bm k};\omega_{m} , {\bm k'})\tau_{z}$ is $ {(2\pi)^{3}(\hbar \beta)(-2\epsilon_{{\bm k}}) \over \hbar^{2}\omega_{n}^{2} + \epsilon_{{\bm k}}^{2} +\Delta^{2}}\delta_{n,m}\delta({\bm k}-{\bm k'})$. For the first order contribution of $G_{\text{int}}^{-1}$, we have the following:
\begin{eqnarray}
\text{tr}(G_{0}G^{-1}_{\text{int}}) &=& {g_{f}\mu_{B}\mu_{0}e \hbar \over m_{e}}\int_{0}^{\beta \hbar } dx_{0}\left( \int_{V_{B}}d^{3}x \overline{{\bm \phi}} \mathuu{\sigma_{i}} {\bm \phi} \right)\epsilon^{ij\ell} \nonumber \\ &{}& \left( {1\over \beta \hbar}\sum_{(m,n) \in \mathbb{Z}\times \mathbb{Z}}\int {d^{3}kd^{3}k' \over (2\pi)^{3}} { \hbar \omega_{n}(k_{\ell}-k_{\ell}')(k_{j}+k_{j}') \over (\hbar^{2}\omega_{n}^{2} + \epsilon_{{\bm k}}^{2} +\Delta^{2})\Vert {\bm k}-{\bm k'} \Vert^{2}}e^{-i({\bm k}-{\bm k'})\cdot {\bm x}}e^{ix_{0}(\omega_{m}-\omega_{n})}\delta({\bm k}-{\bm k'})\delta_{\omega_{n},\omega_{m}} \right. \nonumber \\ &+& \left. i{2m_{e} \over \hbar (\hbar \beta)^{2}(2\pi)^{6}}  \sum_{(m,n,s) \in \mathbb{Z}^{3}}\int d^{3}kd^{3}k'd^{3}q  { \epsilon_{{\bm k}'}(k_{\ell}'-k_{\ell}-q_{\ell})v_{S,j}(\omega_{s},{\bm q})e^{-i ({\bm k'}-{\bm k}-{\bm q})\cdot {\bm x}} \over (\hbar^{2}\omega_{m}^{2} + \epsilon_{{\bm k'}}^{2} +\Delta^{2})\Vert {\bm k'}-{\bm k} -{\bm q}\Vert^{2}} \right. \nonumber \\ &{}& \left. e^{ix_{0}(\omega_{s}-(\omega_{n}-\omega_{m}))}\delta({\bm k}-{\bm k'})\delta_{\omega_{n},\omega_{m}} \right)
\end{eqnarray}
It is clear that the $\delta(k_{\ell}-k'_{\ell})$ integration causes the first term to vanish. That the second term vanishes follows from the fact that \begin{equation}
\int {d^{3}k \over (2\pi)^{3}} {\epsilon_{{\bm k}}\over (\hbar^{2}\omega_{m}^{2} + \epsilon_{{\bm k}}^{2} +\Delta^{2})} ={1\over (2\pi)^{2}} \int_{-\hbar \omega_{D}}^{\hbar \omega_{D}} d\epsilon \rho(\epsilon) {\epsilon \over (\hbar^{2}\omega_{m}^{2} + \epsilon^{2} +\Delta^{2})}  = 0 ,
\end{equation}
where $\rho(\epsilon) := {m_{e}\over \hbar^{2}}\sqrt{2m_{e}\epsilon \over \hbar^{2}}$.

From Eq.\,(\ref{eqn:perturbs}), it is clear that  $G_{\dot{\theta}}^{-1}$ vanishes if we assume that the imaginary time Josephson-Anderson relation (imaginary time a.c. Josephson equation) ${\del \theta_{L(R)} \over \del \tau} = {2e\over \hbar}\varphi_{L(R)}$ holds on the left- and right-hand sides of the Josephson junction, respectively. Hence, the second order cross-term $-{1\over 2}\text{tr}(G_{0} G_{\text{int}}^{-1}G_{0}G_{\dot{\theta}}^{-1})$ is zero. The second-order term $-{1\over 2}\text{tr}(G_{0} G_{\dot{\theta}}^{-1}G_{0}G_{\dot{\theta}}^{-1})$ gives rise to the capacitive term in the flux action Eq.\,(\ref{eqn:fluxactbare}) \cite{ambegaokar}.

The second order cross-term $-{1\over 2}\text{tr}(G_{0}(\omega_{n},{\bm k}) G_{\text{int}}^{-1}(\omega_{n},{\bm k} ; \omega_{m},{\bm k}') G_{0}(\omega_{m},{\bm k'})G_{{\bm v}_{S}}^{-1}(\omega_{m},{\bm k}';\omega_{n},{\bm k}))$ contains one term of order $\mathcal{O}(\Vert {\bm v}_{S}\Vert^{3})$, which we ignore. In addition, if we make the assumption that the supercurrent has zero divergence, then ${\bm k}\cdot {\bm v}_{S} = 0$ so that the first term of $G_{{\bm v}_{S}}^{-1}$ in Eq.\,(\ref{eqn:perturbs}) does not contribute at second order in perturbation theory. The remaining term is proportional to
\begin{eqnarray}
{} &{}& i\int d^{3}kd^{3}k'd^{3}q\,\sum_{(n,m,r) \in \mathbb{Z}^{3}}\int_{0}^{\beta \hbar}dx_{0} \int_{V_{B}}^{({\bm x})} [{\bm M}(x_{0},{\bm x}) \cdot ({\bm k} + {\bm k}') \times ({\bm k} - {\bm k}')] \nonumber \\ &{}& {2i\hbar (\omega_{n} + \omega_{m})\epsilon_{\bm k} \over (\hbar^{2}\omega_{n}^{2} + \epsilon_{\bm k}^{2} - \Delta^{2})(\hbar^{2}\omega_{m}^{2} + \epsilon_{{\bm k}'}^{2} - \Delta^{2})}{e^{ix_{0}(\omega_{m}-\omega_{n})}e^{-i({\bm k} - {\bm k}')\cdot {\bm x}} \over \Vert {\bm k} - {\bm k}' \Vert^{2}}{\bm v}_{S}(\omega_{r},{\bm q})\cdot {\bm v}_{S}(\omega_{m}-\omega_{n}-\omega_{r},{\bm k} - {\bm k}' - {\bm q}) .
\label{eqn:ccc}
\end{eqnarray}
Making the approximation ${\bm k} = {\bm k}'$ due to the fact that ${\bm v_{S}}$ can be considered to have only a single Fourier component, the above integration takes the simplified form $\int d^{3}k \, 2({\bm M}\cdot {\bm k})g(\Vert {\bm k} \Vert)$, where $g$ symbolizes the even parity part of the integrand. This integral evaluates to zero because of the odd parity of ${\bm k}$. 

Finally, we show in detail the derivation of Eq.\,(\ref{eqn:secondordergint}) from Eq.\,(\ref{eqn:intermed}), which involves an approximation to an effective action that is nonlocal in imaginary time. In order to simplify the discussion, we derive $S^{(2)}_{\text{eff,int}}$ in the $T=0$ limit. In Eq.\,(\ref{eqn:intermed}), the factor $S(\omega_{n},\omega_{r},\epsilon)$ is given by
\begin{equation}
{\left( \hbar^{2}\omega_{n}\omega_{r} + \Delta^{2}-\epsilon^{2} \right) e^{i\omega_{n}(x_{0}'-x_{0})}e^{i\omega_{r}(x_{0}-x_{0}')} \over (\hbar^{2}\omega_{n}^{2} + \epsilon^{2} - \Delta^{2})(\hbar^{2}\omega_{r}^{2} + \epsilon^{2} - \Delta^{2}) }.
\end{equation} For $\beta \rightarrow \infty$, the residue theorem is used to evaluate the sum $\sum_{(n,r)\in \mathbb{Z}\times \mathbb{Z}}S(\omega_{n},\omega_{r},\epsilon)$.  The result is:
\begin{equation}
\sum_{(n,r) \in \mathbb{Z} \times \mathbb{Z}}S(\omega_{n},\omega_{r},\epsilon) = {1\over 4\hbar^{2}}\left( 1+{\Delta^{2}-\epsilon^{2} \over \Delta^{2} + \epsilon^{2}}\right) e^{-\sqrt{\epsilon^{2}+\Delta^{2}}\vert x_{0} - x_{0}' \vert \over \hbar} .
\label{eqn:nonlocsupp}
\end{equation}  Taking $x_{0}= x_{0}'$, we have ${1\over 4\hbar^{2}}\int_{-\hbar \omega_{D}}^{\hbar \omega_{D}}d\epsilon \,\left( 1+{\Delta^{2}-\epsilon^{2} \over \Delta^{2} + \epsilon^{2}}\right)= {\Delta \over \hbar^{2}}\tan^{-1}(\hbar \omega_{D} / \Delta) > 0$. Carrying out the $\int_{0}^{\beta \hbar}dx_{0}'$ integration gives approximately $\hbar / \Delta$ because this is the characteristic inverse frequency scale of superconductivity. Equation (\ref{eqn:secondordergint}) then follows from the line of reasoning presented in the main text.

However, it is clear that Eq.\,(\ref{eqn:nonlocsupp}) is nonlocal in imaginary time. We have \begin{eqnarray}
\int_{-\hbar \omega_{D}}^{\hbar \omega_{D}}d\epsilon \, \sum_{(n,r) \in \mathbb{Z} \times \mathbb{Z}}S(\omega_{n},\omega_{r},\epsilon) &=& {1\over 2\hbar^{2}}\int_{-\hbar \omega_{D}}^{\hbar \omega_{D}}d\epsilon \, {\Delta^{2} \over \epsilon^{2} + \Delta^{2}}e^{-\sqrt{\epsilon^{2}+\Delta^{2}}\vert x_{0} - x_{0}' \vert \over \hbar} \nonumber \\ &{=}& {\Delta^{2} \over \hbar^{2}}\int_{0}^{\sinh^{-1}(\hbar \omega_{D}/\Delta)}du(\Delta \cosh u)^{-1}e^{-\Delta \vert x_{0} - x_{0}'\vert \cosh(u)/\hbar}.
\end{eqnarray}
\end{widetext}
Using the $T=0$ BCS gap equation $V_{0}\rho(E_{F}) = \sinh^{-1}(\hbar \omega_{D}/\Delta)$ where $V_{0}$ is the effective electron-electron attraction, and taking $V_{0}\rho(E_{F})$ to be large, the integral above is approximately ${1\over \hbar^{2}}\int_{\Delta \vert x_{0} - x_{0}'\vert /\hbar}^{\infty} dx K_{0}(x) =: Ki_{1}(\Delta \vert x_{0} - x_{0}'\vert /\hbar)$, where $Ki_{1}(x)$ is the Bickley function \cite{bickleynayler}. This nonlocal kernel can be used to carry out an analysis of the dissipative evolution of the hybrid BEC--SFQ system.

\bibliography{am22.bib}

\end{document}